\documentclass[pra,twocolumn,showpacs,preprintnumbers,superscriptaddress]
{revtex4}

\usepackage{times}
\usepackage{bm}
\usepackage{float}
\usepackage{graphicx}
\usepackage{amsbsy}
\usepackage{amsmath}
\usepackage{amsfonts}
\usepackage{amsthm}
\begin{document}

\theoremstyle{plain}
\newtheorem{theorem}{Theorem}
\newtheorem{lemma}[theorem]{Lemma}
\newtheorem{corollary}[theorem]{Corollary}
\newtheorem{proposition}[theorem]{Proposition}
\newtheorem{conjecture}[theorem]{Conjecture}

\theoremstyle{definition}
\newtheorem{definition}[theorem]{Definition}

\theoremstyle{remark}
\newtheorem*{remark}{Remark}
\newtheorem{example}{Example}
\title{Detection of mixed bipartite entangled state in arbitrary dimension via structural physical approximation of partial transposition}
\author{Anu Kumari, Satyabrata Adhikari}
\email{mkumari_phd2k18@dtu.ac.in, satyabrata@dtu.ac.in} \affiliation{Delhi Technological
University, Delhi-110042, Delhi, India}

\begin{abstract}
It is very crucial to know that whether the quantum state generated in the experiment is entangled or not. In the literature, this topic was studied extensively and researchers proposed different approaches for the detection of mixed bipartite entangled state in arbitrary dimension. Proceeding in this line of research, we also propose three different criteria for the detection of mixed bipartite negative partial transpose (NPT) entangled state in arbitrary dimension. Our criteria is based on the method of structural physical approximation (SPA) of partial transposition (PT). We have shown that the proposed criteria for the detection of NPT entangled state can be realized experimentally. Two of the proposed  criteria is given in terms of the concurrence of the given state in arbitrary dimension so it is essential to find out the concurrence. Thus, we provide new lower and upper bound of concurrence of the quantum state under investigation in terms of average fidelity of two quantum states and hence these bounds can be realized experimentally. Moreover, we have shown how to perform SPA map on qutrit-qubit system and then explicitly calculated the matrix elements of the density matrix describing the SPA-PT of the qutrit-qubit system. We then illustrate our criteria for the detection of entanglement by considering a class of qubit-qubit system and a class of qutrit-qubit system.
\end{abstract}
 \maketitle

\section{Introduction}
Entanglement \cite{piani} is a basic ingredient that enhanced the power of quantum computation \cite{jozsa}. This feature of quantum mechanics also plays a very important role in the burgeoning field of quantum information theory. Many applications such as quantum teleportation \cite{Bennett1}, superdense coding \cite{Bennett2}, remote state preparation \cite{pati}, quantum cryptography \cite{Gisin} etc., which are now a reality, based on the idea of quantum entanglement. Therefore, it is very important to probe whether the given quantum state is entangled or not? This question is important in the sense that when an experimental set up aims to create an entangled state, it produces a mixed state due to noisy environment but it is not guaranteed that the produced state is entangled. So we need some detection criteria by which we can determine whether the state produced in an experiment is entangled or not. This type of problem is known as entanglement detection problem \cite{guhne2}. Several criteria and many methods had already been proposed for solving the entanglement detection problem but yet partial result has ben obtained (\cite{augusiak}-\cite{toth2}).\\
The first and simplest criteria to detect entanglement is the PT criterion introduced by Peres \cite{peres} and later Horodecki's \cite{horodecki2} proved that it is necessary and sufficient condition for $2\otimes2$ and $2\otimes3$ quantum system. The PT criterion will become only necessary for $d_{1}\otimes d_{2}$ ($d_{1},d_{2}\geq 3$) dimensional quantum system due to the existence of positive partial transpose entangled states (PPTES) in higher dimensional system. Since PPTES are not detected by PT criterion so some other methods are developed to detect PPTES \cite{oliver}. The other weakness of the PT criterion is that it cannot be implemented in the laboratory and this is due to the fact that it is a positive but not a completely positive map. We can overcome this defect by using structural physical approximation of partial transposition (SPA-PT) map. In SPA-PT method, a positive maps can be approximated by completely positive maps and thus transform non-physical partial transposition operations to a physical operations \cite{horodecki3}.\\
Let us consider a $d_{1}\otimes d_{2}$  dimensional quantum system and performing the partial transposition operation on the second subsystem. Since the partial transposition map $id\otimes\textbf{\textsl{T}}$, where $\textbf{\textsl{T}}$ denotes the positive transposition map and $id$ represent the identity operator, is not a physical map so we consider the approximated map $\widetilde{id\otimes\textbf{\textsl{T}}}$, which  is a
completely positive map corresponds to a quantum channel that can be experimentally implementable \cite{bae}. The approximated map can be expressed as \cite{horodecki3}
\begin{eqnarray}
\widetilde{id\otimes\textbf{\textsl{T}}}=(1-q^{*})(id\otimes\textbf{\textsl{T}})+\frac{q^{*}}{d_{1}d_{2}}I_{A}\otimes
I_{B} \label{spapt}
\end{eqnarray}
where $q^{*}$ denote the minimum value of $q$ for which $\widetilde{id\otimes\textbf{\textsl{T}}}$ become a positive semi-definite operator. Since $\widetilde{id\otimes\textbf{\textsl{T}}}$ is a completely positive map so when it operate on a density matrix $\rho^{in}$, it gives another density matrix $\widetilde{\rho}^{out}$ at the output, i.e. $(\widetilde{id\otimes\textbf{\textsl{T}}})\rho^{in}=\widetilde{\rho}^{out}$. The minimum eigenvalue of the density matrix $\widetilde{\rho}^{out}$ is important in the sense that there exist a critical value of the minimum eigenvalue $\lambda_{min}(\widetilde{\rho}^{out})$ below which the state described by the density matrix $\rho^{in}$ is entangled. For instance, $2\otimes 2$ dimensional quantum state is entangled when the minimum eigenvalue $\lambda_{min}(\widetilde{\rho}^{out})$ is less than $\frac{2}{9}$. So the analysis of minimum eigenvalue of $\widetilde{\rho}^{out}$ is needed. It has been found that SPA-PT method not only detect entanglement but also can be applied to estimate the optimal singlet fraction, entanglement negativity and concurrence of the two-qubit density matrix in an experiment \cite{adhikari1,adhikari2}. H. T. Lim et.al. have studied the SPA-PT of the qutrit and shown that it can be implemented in the laboratory with linear optical elements \cite{lim1}.\\
Since SPA-PT method works well in two qubit system and can be implemented in an experiment so it is important to ask if we can use this method to detect entanglement in higher dimensional system. The answer is in affirmative. In this work, we consider $d_{1}\otimes d_{2}$ dimensional quantum system and derived three different criteria for the detection of negative partial transpose entangled state (NPTES) employing SPA-PT method. It is known that the average fidelity between two quantum states can be realized in experiment \cite{kwong}. Thus, we have expressed our proposed criteria to detect NPTES, in terms of the average fidelity between two quantum states and hence the detection criteria can be implemented in an experiment. Among three criteria, two of them is given in terms of the concurrence of the given state so it is essential to find out the concurrence. As far as we know, there does not exist any procedure to calculate the actual value of the concurrence of the given state in arbitrary dimension but in the literature, there exist lower bound of the concurrence \cite{mintert1}. In the present work also, we have provided a lower and upper bound of the concurrence that can be realized in an experiment. Secondly, for $d\otimes d$ dimensional quantum system, the minimum eigenvalue $\lambda_{min}(\widetilde{\rho}^{out})$ is $\frac{d^{2}\mu}{d^{4}\mu+1}$, where $\mu=-\mu'$, $\mu'$ is the most negative eigenvalue of $id\otimes\textbf{\textsl{T}}$ \cite{horodecki3}. Since $id\otimes\textbf{\textsl{T}}$ is not a physical operation so its minimum eigenvalue cannot be determined in experiment. Thus, our idea here is to determine the density matrix of SPA-PT of any arbitrary dimensional quantum system and hence once density matrix is in our possession, one can calculate the minimum eigenvalue that can be used in the detection of entanglement. Specifically, we have operated SPA map on qutrit-qubit system and then explicitly constructed the output density matrix describing the SPA-PT of qutrit-qubit system.\\
This paper is organized as follows: In Sec. II, we have stated some preliminary result that will be needed in the later section.
In Sec. III, we have derived the criteria for the detection of NPTES in arbitrary dimensional quantum system. In Sec. IV, we have shown how to apply our proposed detection criteria by considering qubit-qubit and qutrit-qubit quantum system. We conclude in Sec. V.

\section{Preliminary Results}
\textbf{Result-1:}
For any two hermitian (n,n) matrices A and B, we have
\begin{eqnarray}
\lambda_{min}(A)Tr(B)\leq Tr(AB)\leq \lambda_{max}(A)Tr(B)
\label{result1}
\end{eqnarray}
\textbf{Proof:} It is known that for any hermitian (n,n) matrices A and B, the following inequality holds \cite{Lasserre}
\begin{eqnarray}
\sum_{i=1}^{n}\lambda_{i}(A)\lambda_{n-i+1}(B)\leq Tr(AB)\leq \sum_{i=1}^{n}\lambda_{i}(A)\lambda_{i}(B)
\label{theorem1}
\end{eqnarray}
where $\lambda_{min}=\lambda_{1}\leq \lambda_{2}\leq \lambda_{3}\leq.......\leq \lambda_{n}=\lambda_{max}$.\\
In the LHS of inequality (\ref{theorem1}), if we replacing all eigenvalues of A by its minimum eigenvalue and in the RHS, if we replace all eigenvalues of A by its maximum, then the inequality (\ref{theorem1}) reduces to (\ref{result1}). Hence proved.\\
\textbf{Result-2:} If $W$ represent the witness operator that detect the entangled quantum state described by the density operator $\rho_{AB}$ and
$C(\rho_{AB})$ denote the concurrence of the state $\rho_{AB}$ then the lower bound of concurrence is given by \cite{mintert2}
\begin{eqnarray}
C(\rho_{AB})\geq -Tr[W\rho_{AB}]
\label{result2}
\end{eqnarray}
\textbf{Result-3:} If $\rho_{AB}$ denote the density operator of a bipartite quantum state in any arbitrary dimension then all eigenvalues of $\rho_{AB}^{T_{B}}$ lying within the interval $[\frac{-1}{2},1]$ \cite{rana}.\\
\textbf{Result-4:} If any arbitrary two qubit density operator $\rho_{12}$ is given by
\begin{eqnarray}
\rho_{12}=
\begin{pmatrix}
  e_{11} & e_{12} & e_{13} & e_{14} \\
  e_{12}^{*} & e_{22} & e_{23} & e_{24} \\
  e_{13}^{*} & e_{23}^{*} & e_{33} & e_{34} \\
  e_{14}^{*} & e_{24}^{*} & e_{34}^{*} & e_{44}
\end{pmatrix}, \sum_{i=1}^{4}e_{ii}=1
\end{eqnarray}
where $(*)$ denotes the complex conjugate, then the structural physical approximation of partial transpose of $\rho_{12}$ is given by \cite{adhikari1}
\begin{eqnarray}
\widetilde{\rho_{12}}&=&[\frac{1}{3}(I\otimes\widetilde{T})+\frac{2}{3}(\widetilde{\Theta}\otimes\widetilde{D})]\rho_{12}\nonumber\\&=&
\begin{pmatrix}
  E_{11} & E_{12} & E_{13} & E_{14} \\
  E_{12}^{*} & E_{22} & E_{23} & E_{24} \\
  E_{13}^{*} & E_{23}^{*} & E_{33} & E_{34} \\
  E_{14}^{*} & E_{24}^{*} & E_{34}^{*} & E_{44}
\end{pmatrix}
\label{spa1}
\end{eqnarray}
where
\begin{eqnarray}
&&E_{11}=\frac{1}{9}(2+e_{11}),E_{12}=\frac{1}{9}e_{12}^{*}, E_{13}=\frac{1}{9}e_{13},\nonumber\\&&
E_{14}=\frac{1}{9}e_{23}, E_{22}=\frac{1}{9}(2+e_{22}),E_{23}=\frac{1}{9}e_{14},\nonumber\\&&
E_{24}=\frac{1}{9}e_{24},E_{33}=\frac{1}{9}(2+e_{33}),E_{34}=\frac{1}{9}e_{34}^{*},\nonumber\\&&
E_{44}=\frac{1}{9}(2+e_{44})
\label{spa2a}
\end{eqnarray}
\section{Criteria for the detection of bipartite Negative Partial Transpose Entangled states in arbitrary dimension}
In this section, we derive the criteria for the detection of bipartite negative partial transpose entangled states in arbitrary dimension employing the method of structural physical approximation (SPA) of partial transposition of any arbitrary dimensional quantum state.
\subsection{Lower and Upper bound of the minimum eigenvalue of SPA of partial transpose of a bipartite mixed quantum state in arbitrary dimension}
Let us consider a bipartite mixed quantum state in arbitrary dimension described by the density operator $\rho_{AB}$. If $\tilde{\rho}_{AB}$ denote the SPA of the partial transpose of $\rho_{AB}$ and $Q$ be any positive semi-definite operator such that $Tr(Q)=1$ then the quantity $Tr[(\tilde{\rho}_{AB}+Q^{T_B})\rho_{AB}]$, where $T_{B}$ denote the partial transposition with respect to the system B, may be expressed as
\begin{eqnarray}
Tr[(\tilde{\rho}_{AB}+Q^{T_B})\rho_{AB}]&=&Tr[\tilde{\rho}_{AB}\rho_{AB}+Q^{T_B}\rho_{AB}]\nonumber\\&=&Tr[\tilde{\rho}_{AB}\rho_{AB}]+Tr[Q^{T_B}\rho_{AB}]
\nonumber\\&=&Tr[\tilde{\rho}_{AB}\rho_{AB}]+Tr[Q\rho_{AB}^{T_B}]
\label{quantity}
\end{eqnarray}
Taking $A=\tilde{\rho}_{AB}$ and $B=\rho_{AB}$ in the LHS part of (\ref{result1}), we get
\begin{eqnarray}
\lambda_{min}(\tilde{\rho}_{AB})\leq Tr(\tilde{\rho}_{AB}\rho_{AB})
\label{ineq1}
\end{eqnarray}
Similarly, applying result-1 with $A=\rho_{AB}^{T_B}$ and $B=Q$, we obtain
\begin{eqnarray}
\lambda_{min}(\rho_{AB}^{T_B})\leq Tr[Q\rho_{AB}^{T_B}]=Tr[Q^{T_{B}}\rho_{AB}]
\label{ineq2}
\end{eqnarray}
Adding (\ref{ineq1}) and (\ref{ineq2}), we get
\begin{eqnarray}
&&\lambda_{min}(\tilde{\rho}_{AB})+\lambda_{min}(\rho_{AB}^{T_B})\leq Tr[\tilde{\rho}_{AB}\rho_{AB}]+Tr[Q^{T_{B}}\rho_{AB}]\nonumber\\&&
\Rightarrow \lambda_{min}(\rho_{AB}^{T_B})\leq G
\label{ineq3}
\end{eqnarray}
where $G=Tr[\tilde{\rho}_{AB}\rho_{AB}]+Tr[Q^{T_{B}}\rho_{AB}]-\lambda_{min}(\tilde{\rho}_{AB})$.\\
Therefore, the above inequality gives the upper bound on minimum eigenvalue of $\rho_{AB}^{T_B}$.\\
A bipartite density operator $\rho_{AB}$ in any arbitrary dimension, represent an NPT entangled state if and only if $\lambda_{min}(\rho_{AB}^{T_B})$ is negative. Further, if we assume that NPT entangled state described by the density operator $\rho_{AB}$
detected by the witness operator $W=Q^{T_{B}}$ then we have
\begin{eqnarray}
&&\lambda_{min}(\tilde{\rho}_{AB})\geq Tr[\tilde{\rho}_{AB}\rho_{AB}]+Tr[W\rho_{AB}]
\label{lb}
\end{eqnarray}
Again, inequality (\ref{ineq3}) can be re-expressed as
\begin{eqnarray}
\lambda_{min}(\tilde{\rho}_{AB}) &\leq& -\lambda_{min}(\rho_{AB}^{T_B})+Tr[\tilde{\rho}_{AB}\rho_{AB}]\nonumber\\&+&Tr[Q^{T_{B}}\rho_{AB}]
\label{ineq4}
\end{eqnarray}
Using Result-3 in (\ref{ineq4}) and $W=Q^{T_{B}}$, the inequality (\ref{ineq4}) reduces to
\begin{eqnarray}
\lambda_{min}(\tilde{\rho}_{AB})\leq \frac{1}{2}+Tr[\tilde{\rho}_{AB}\rho_{AB}]+Tr[W\rho_{AB}]
\label{ub}
\end{eqnarray}
Combining inequalities (\ref{lb}) and (\ref{ub}), we get the lower bound (L) and upper bound (U) of minimum eigenvalue of the SPA of partial transpose of $\rho_{AB}^{T_{B}}$ and they are given by
\begin{eqnarray}
L \leq \lambda_{min}(\tilde{\rho}_{AB}) \leq U
\label{ub1}
\end{eqnarray}
where
\begin{eqnarray}
L= Tr[\tilde{\rho}_{AB}\rho_{AB}]+Tr[W\rho_{AB}]
\label{lu}
\end{eqnarray}
\begin{eqnarray}
U= \frac{1}{2}+Tr[\tilde{\rho}_{AB}\rho_{AB}]+Tr[W\rho_{AB}]
\label{ub100}
\end{eqnarray}
We note that the lower bound given by $L$ can be negative also but since the minimum eigenvalue
$\lambda_{min}(\tilde{\rho}_{AB})$ of the positive semi-definite operator $\tilde{\rho}_{AB}$ is always positive
so the inequality (\ref{ub1}) can be re-expressed as
\begin{eqnarray}
max\{L,0\} \leq \lambda_{min}(\tilde{\rho}_{AB}) \leq U
\label{bound}
\end{eqnarray}
where $L$ and $U$ are given by (\ref{lu}) and (\ref{ub100}) respectively.
\subsection{Criteria for the detection of $d_{1}\otimes d_{2}$ dimensional bipartite NPT entangled state}
The entanglement of a bipartite mixed quantum state $\rho_{AB}$ in $d_{1}\otimes d_{2}$ dimension can be detected by computing the value of $L$. To see this,
let us consider the second term of $L$, which is given by $Tr(W\rho_{AB})=Tr(Q^{T_{B}}\rho_{AB})$. Since the witness operator has been constructed
by taking the partial transpose of a positive semi-definite operator so it is not physically realizable and so we would like to approximate
the witness operator $W$ in such a way that it would become a completely positive operator. Therefore, if $\widetilde{W}$ be the approximation of the witness operator $W$ then it can be expressed as
\begin{eqnarray}
\widetilde{W}=pW+\frac{1-p}{d_{1}d_{2}}I, 0\leq p\leq1,
\label{approxw}
\end{eqnarray}
The value of the parameter $p$ should be chosen in such a way that $\widetilde{W}$ become a positive semi-definite operator. Further, we note
that $Tr(\widetilde{W})=1$. Thus, the operator $\widetilde{W}$ represent a quantum state.\\
\textbf{Criterion-1:} A bipartite $d_{1}\otimes d_{2}$ dimensional quantum state described by the density operator $\rho_{AB}$ represent a NPT entangled state iff
\begin{eqnarray}
Tr(\widetilde{W}\rho_{AB})<\frac{1-p}{d_{1}d_{2}}=R
\label{criteria1}
\end{eqnarray}
\textbf{Proof:} Using (\ref{approxw}), the relation between $Tr(W\rho)$ and $Tr(\widetilde{W}\rho)$ can be established as
\begin{eqnarray}
Tr(W\rho_{AB})=\frac{1}{p}[Tr(\widetilde{W}\rho_{AB})-\frac{1-p}{d_{1}d_{2}}]
\label{expectation1}
\end{eqnarray}
Since $W$ denote the witness operator that detect the NPT entangled state $\rho_{AB}$ so $Tr(W\rho_{AB})<0$
and hence proved the required criterion.\\
Since $\widetilde{W}$ have all the properties of a quantum state so $Tr(\widetilde{W}\rho_{AB})$ can be considered same as
the average fidelity between two mixed quantum state $\widetilde{W}$ and $\rho_{AB}$ and therefore, it is given by \cite{kwong}
\begin{eqnarray}
Tr(\widetilde{W}\rho_{AB})=F_{avg}(\widetilde{W},\rho_{AB})
\label{averagefidelity}
\end{eqnarray}
It is also known that the average fidelity $F_{avg}(\widetilde{W},\rho_{AB})$
can be estimated experimentally by Hong-Ou-Mandel interferometry and thus equation (\ref{criteria1}) can be re-expressed
as
\begin{eqnarray}
F_{avg}(\widetilde{W},\rho_{AB})<\frac{1-p}{d_{1}d_{2}}
\label{criteria1mod}
\end{eqnarray}
\textbf{Criterion-2:} The bipartite state $\rho_{AB}$ of any arbitrary dimension is NPT entangled iff
 \begin{eqnarray}
\lambda_{min}(\tilde{\rho}_{AB})\geq F_{avg}(\rho_{AB},\tilde{\rho}_{AB})-C(\rho_{AB})
\label{entcrit2}
\end{eqnarray}
where $C(\rho_{AB})$ denote the concurrence of the density operator $\rho_{AB}$. The lower and upper bound of $C(\rho_{AB})$ is given by
\begin{eqnarray}
\frac{1-p}{pd_{1}d_{2}}-\frac{F_{avg}(\widetilde{W},\rho_{AB})}{p}\leq C(\rho_{AB})\leq F_{avg}(\rho_{AB},\tilde{\rho}_{AB})
\label{concbounded}
\end{eqnarray}
\textbf{Proof:} Let us first recall the lower bound $L$ of the minimum eigenvalue $\lambda_{min}(\tilde{\rho}_{AB})$ and using Result-2 in (\ref{lu}), we get
\begin{eqnarray}
L \geq Tr[\tilde{\rho}_{AB}\rho_{AB}]-C(\rho_{AB})
\label{lbconc}
\end{eqnarray}
The RHS of the inequality (\ref{lbconc}) may take non-negative or negative value. Consider the case
when $Tr[\tilde{\rho}_{AB}\rho_{AB}]-C(\rho_{AB})$ is non-negative i.e.
\begin{eqnarray}
C(\rho_{AB})\leq Tr[\tilde{\rho}_{AB}\rho_{AB}]=F_{avg}(\rho_{AB},\tilde{\rho}_{AB})
\label{conc100}
\end{eqnarray}
Then using (\ref{lbconc}), we can re-write (\ref{lb}) as
\begin{eqnarray}
\lambda_{min}(\tilde{\rho}_{AB})\geq F_{avg}(\rho_{AB},\tilde{\rho}_{AB})-C(\rho_{AB})
\label{entcrit200}
\end{eqnarray}
Moreover, using Result-2, (\ref{expectation1}) and (\ref{averagefidelity}), we get
\begin{eqnarray}
C(\rho_{AB})\geq \frac{1}{p}[\frac{1-p}{d_{1}d_{2}}-F_{avg}(\widetilde{W},\rho_{AB})]
\label{concurrence1}
\end{eqnarray}
Combining (\ref{conc100}) and (\ref{concurrence1}), we get
\begin{eqnarray}
\frac{1-p}{pd_{1}d_{2}}-\frac{F_{avg}(\widetilde{W},\rho_{AB})}{p}\leq C(\rho_{AB})\leq F_{avg}(\rho_{AB},\tilde{\rho_{AB}})
\label{concbounded}
\end{eqnarray}
Hence proved.\\
Since the average fidelity $F_{avg}(\widetilde{W},\rho_{AB})$ and $F_{avg}(\rho_{AB},\tilde{\rho}_{AB})$
can be estimated experimentally so the lower bound and upper bound of $C(\rho_{AB})$ can be estimated experimentally.\\
\textbf{Criterion-3:} The bipartite state $\rho_{AB}$ of any arbitrary dimension is NPT entangled iff
 \begin{eqnarray}
U^{ent}<\frac{1}{2}
\label{entcrit3}
\end{eqnarray}
where $U^{ent}=\frac{1}{2}+F_{avg}(\rho_{AB},\tilde{\rho_{AB}})-C(\rho_{AB})$ and $C(\rho_{AB})$ is given by
\begin{eqnarray}
C(\rho_{AB})> F_{avg}(\rho_{AB},\tilde{\rho_{AB}})
\label{concbounded100}
\end{eqnarray}
\textbf{Proof:} Let us recall the upper bound $U$ of the minimum eigenvalue $\lambda_{min}(\tilde{\rho}_{AB})$ and using Result-2 in (\ref{ub100}), we get
\begin{eqnarray}
U \geq \frac{1}{2}+ Tr[\tilde{\rho}_{AB}\rho_{AB}]-C(\rho_{AB})=U^{ent}
\label{ubconc}
\end{eqnarray}
If the quantity $Tr[\tilde{\rho}_{AB}\rho_{AB}]-C(\rho_{AB})$ is non-negative then $U \geq \frac{1}{2}$. Also if the state is separable i.e. if $C(\rho_{AB})=0$ then the value of $U$ again comes out to be greater than $\frac{1}{2}$. So it would be difficult to detect
the entangled state when $Tr[\tilde{\rho}_{AB}\rho_{AB}]-C(\rho_{AB})\geq0$. If we now consider the case when $Tr[\tilde{\rho}_{AB}\rho_{AB}]-C(\rho_{AB})<0$, which indeed may be the case, then we can obtain
$U^{ent}<\frac{1}{2}$. Thus, we can infer when $U^{ent}<\frac{1}{2}$ for $C(\rho_{AB})>Tr[\tilde{\rho}_{AB}\rho_{AB}]=F_{avg}(\rho_{AB},\tilde{\rho_{AB}})$, the state is NPT entangled state. Hence the criterion.
\section{Illustrations}
In this section, we will verify our entanglement detection criteria given in the previous section by taking the example of a class of qubit-qubit system and qutrit-qubit system.
\subsection{Qubit-Qubit system}
Let us consider a large class of two qubit system described by the density operator $\rho_{AB}$ given as \cite{connor}
\begin{eqnarray}
\rho^{(1)}_{AB}=
\begin{pmatrix}
  a & 0 & 0 & 0 \\
  0 & b & f & 0 \\
  0 & f^{*} & b & 0 \\
  0 & 0 & 0 & a
\end{pmatrix}, a+b=\frac{1}{2}
\label{twoqubitexample1}
\end{eqnarray}
where $(*)$ denotes the complex conjugate.\\
The two-qubit density matrix of the form (\ref{twoqubitexample1}) has been studied by many authors \cite{bruss}-\cite{munro} as these form of density marix having high entanglement. In particular, Ishizaka and Hiroshima \cite{ishizaka} have studied such density matrix and maximize the entanglement for a fixed set of eigenvalues with one of the eigenvalue is zero.\\
The density operator $\rho_{AB}^{(1)}$ is positive semi-definite only when
\begin{eqnarray}
|f| \leq b
\label{psd}
\end{eqnarray}
The partial transpose of $\rho^{(1)}_{AB}$ is given by
\begin{eqnarray}
(\rho^{(1)}_{AB})^{T_B}=
\begin{pmatrix}
  a & 0 & 0 & f \\
  0 & b & 0 & 0 \\
  0 & 0 & b & 0 \\
  f^{*} & 0 & 0 & a
\end{pmatrix}
\end{eqnarray}
$(\rho^{(1)}_{AB})^{T_B}$ has negative eigenvalue if
\begin{eqnarray}
|f|>a
\label{nev}
\end{eqnarray}
Therefore, the state $\rho^{(1)}_{AB}$ is an entangled state iff
\begin{eqnarray}
a<|f|\leq b~~~~ \textrm{and}~~~~ a+b=\frac{1}{2}
\label{entstate}
\end{eqnarray}
The concurrence of the state $\rho^{(1)}_{AB}$ is given by
\begin{eqnarray}
C(\rho^{(1)}_{AB})=|f|-a
\label{entstate}
\end{eqnarray}
Let us now assume that if $\rho^{(1)}_{AB}$ is an entangled state then it is detected by the witness operator $W^{(1)}$. The witness operator $W^{(1)}$ can be expressed as
\begin{eqnarray}
W^{(1)}=|\psi\rangle^{T_{B}}\langle\psi|,~|\psi\rangle=\frac{1}{\sqrt{1+|k|^{2}}}(k|00\rangle+|11\rangle)
\label{witness1}
\end{eqnarray}
where $k=\frac{-f}{|f|}$.\\
Since it is not possible to implement partial transposition operation experimentally so we use Result-4 to obtain the SPA-PT of $\rho^{(1)}_{AB}$ and $W^{(1)}$. SPA-PT of $\rho^{(1)}_{AB}$ and $W^{(1)}$ are given in the following form
\begin{eqnarray}
\tilde{\rho}^{(1)}_{AB}=
\begin{pmatrix}
 \frac{2+a}{9} & 0 & 0 & \frac{f}{9}\\
 0 & \frac{2+b}{9} & 0 & 0\\
 0 & 0 & \frac{2+b}{9} & 0\\
 \frac{f^{*}}{9} & 0 & 0 & \frac{2+a}{9}
\end{pmatrix}
\end{eqnarray}
\begin{eqnarray}
\tilde{W}^{(1)}=
\begin{pmatrix}
\frac{1}{3} & 0 & 0 & 0\\
0 & \frac{1}{6} & \frac{k}{6} & 0\\
0 & \frac{k^{*}}{6} & \frac{1}{6} & 0\\
0 & 0 & 0 & \frac{1}{3}
\end{pmatrix}
\end{eqnarray}
Also, the average fidelities between two pair of mixed quantum state $(\rho^{(1)}_{AB}, \tilde{\rho}^{(1)}_{AB})$ and $(\tilde{W}^{(1)},\rho^{(1)}_{AB})$ respectively, are given by
\begin{eqnarray}
F_{avg}(\rho_{AB},\tilde{\rho}^{(1)}_{AB})= \frac{2a}{9}(2+a)+\frac{2b}{9}(2+b)
\label{avfidstate}
\end{eqnarray}
\begin{eqnarray}
F_{avg}(\tilde{W}^{(1)},\rho^{(1)}_{AB})=\frac{2a+b-|f|}{3}
\label{avstatewitness}
\end{eqnarray}
Now we are in a position to discuss criterion-1, criterion-2 and criterion-3 for a large class of qubit-qubit system describd by the
density operator $\rho^{(1)}_{AB}$.\\
Criterion-1 for the density operator $\rho^{(1)}_{AB}$ takes the form as
\begin{eqnarray}
&&F_{avg}(\tilde{W}^{(1)},\rho^{(1)}_{AB})<\frac{1}{6}
\label{avstatewitness1}
\end{eqnarray}
The satisfaction of the above criterion is given in Table-I.
\begin{table}
\begin{center}
\caption{Table varifying (\ref{avstatewitness1}) for differnt values of the state parameters}
\begin{tabular}{|c|c|c|c|}\hline
State parameter & $F_{avg}(\tilde{W}^{(1)},\rho^{(1)}_{AB})$ & Criterion-1 & Nature of $\rho^{(1)}_{AB})$ \\ (a, b, f) & & & \\  \hline
(0.05, 0.45, 0.4+0.1i)  & 0.04589 & satisfied & Entangled\\\hline
(0.1, 0.4, 0.25+0.25i)  & 0.08214 & satisfied & Entangled\\\hline
(0.15, 0.35, 0.24+0.2i) & 0.11253 & satisfied & Entangled\\\hline
(0.2, 0.3, 0.27+0.13i) & 0.13344 & satisfied & Entangled\\\hline
\end{tabular}
\end{center}
\end{table}
Next, let us illustrate criterion-2 for the density operator $\rho^{(1)}_{AB}$. Criterion-2 can be re-written for $\rho^{(1)}_{AB}$ as
\begin{eqnarray}
\lambda_{min}(\tilde{\rho}^{(1)}_{AB})\geq F_{avg}(\rho^{(1)}_{AB},\tilde{\rho}^{(1)}_{AB})-C(\rho^{(1)}_{AB})
\label{entcrit2ex}
\end{eqnarray}
where $C(\rho_{AB})$ is given by
\begin{eqnarray}
\frac{1}{2}-3F_{avg}(\widetilde{W}^{(1)},\rho^{(1)}_{AB})\leq C(\rho^{(1)}_{AB})\leq F_{avg}(\rho^{(1)}_{AB},\tilde{\rho}^{(1)}_{AB})
\label{concboundedex}
\end{eqnarray}
We can now verify (\ref{entcrit2ex}) and (\ref{concboundedex}) by taking different values of the parameters of the state $\rho^{(1)}_{AB}$. The newly derived lower and upper bound of the concurrence $C(\rho^{(1)}_{AB})$ and the criterion-2 can be verified by Table-II and Table-III respectively.\\
\begin{table}
\begin{center}
\caption{Table varifying (\ref{concboundedex}) for differnt values of the state parameters}
\begin{tabular}{|c|c|c|c|}\hline
State parameter & $F_{avg}(\tilde{W}^{(1)},\rho^{(1)}_{AB})$ & $F_{avg}(\widetilde{\rho}^{(1)}_{AB},\rho^{(1)}_{AB})$ & C($\rho^{(1)}_{AB})$) \\ (a, b, f) & & & \\  \hline
(0.05, 0.45, 0.2+0.2i)  & 0.08905 & 0.26777 & 0.23284\\\hline
(0.1, 0.4, 0.25+0.25i)  & 0.08215 & 0.26 & 0.25355\\\hline
(0.15, 0.35, 0.24+0.2i) & 0.11253 & 0.25444 & 0.16241\\\hline
(0.2, 0.3, 0.27+0.13i) & 0.13344 & 0.25111 & 0.09966\\\hline
\end{tabular}
\end{center}
\end{table}
\begin{table}
\begin{center}
\caption{Table varifying (\ref{entcrit2ex}) for differnt values of the state parameters}
\begin{tabular}{|c|c|c|c|}\hline
State parameter & $\lambda_{min}(\tilde{\rho}^{(1)}_{AB})$ & Criterion-2 & Nature of $\rho^{(1)}_{AB}$   \\ (a, b, f) & & & \\  \hline
(0.05, 0.45, 0.2+0.2i)  & 0.19635 & satisfied & Entangled\\\hline
(0.1, 0.4, 0.25+0.25i)  & 0.19405 & satisfied & Entangled\\\hline
(0.15, 0.35, 0.24+0.2i) & 0.20417 & satisfied & Entangled\\\hline
(0.2, 0.3, 0.27+0.13i) & 0.21114 & satisfied & Entangled\\\hline
\end{tabular}
\end{center}
\end{table}

\subsection{Qutrit-Qubit system}
It is known that certain qutrit-qubit entangled state shows the property of time-invariant entanglement under collective dephasing. Further,
it has been observed that all dimension of Hilbert space studied so far does not exhibit simultaneously two important properties such as
time-invariant entanglement and freezing dynamics. The qutrit-qubit system given in the example below is important in the sense that it exhibit time-invariant entanglement as well as freezing dynamics of entanglement under collective dephasing \cite{ali,karpat}.\\
Let us now consider an example of a qutrit-qubit quantum state described by the density operator $\rho^{(2)}_{AB}$,
\begin{eqnarray}
\rho^{(2)}_{AB}=
\begin{pmatrix}
  0 & 0 & 0 & 0 & 0 & 0\\
  0 & \frac{\alpha}{2} & 0 & 0 & \frac{\alpha}{2} & 0 \\
  0 & 0 & \frac{1-\alpha}{2} & 0 & 0 & \frac{1-\alpha}{2} \\
  0 & 0 & 0 & 0 & 0 & 0\\
  0 & \frac{\alpha}{2} & 0 & 0 & \frac{\alpha}{2} & 0 \\
  0 & 0 & \frac{1-\alpha}{2} & 0 & 0 & \frac{1-\alpha}{2} \\
\end{pmatrix}, 0 \leq \alpha \leq 1
\end{eqnarray}
It has been found that the state $\rho^{(2)}_{AB}$ is entangled for $0\leq \alpha \leq 1$ \cite{ali}. This result can be verified by
the witness operator method. The witness operator that detect the quantum state $\rho^{(2)}_{AB}$ as an entangled state is given by
\begin{eqnarray}
W^{(2)}=|\chi\rangle^{T_{B}}\langle\chi|,~|\chi\rangle=\frac{1}{\sqrt{1+|\kappa|^{2}}}(-\kappa|11\rangle+|20\rangle)
\label{witness2}
\end{eqnarray}
where $\kappa=\frac{\alpha+\sqrt{4-8\alpha+5\alpha^{2}}}{2(1-\alpha)}$.\\
The operator $W^{(2)}$ has been constructed based on the idea of partial transposition but since it is not practically
implementable operation so we use our criteria for the detection of entanglement, which can be implemented in the laboratory.\\
To implement our first criteria, we will use the SPA-PT of $W^{(2)}$, which is given by
\begin{eqnarray}
\tilde{W}^{(2)}=
\begin{pmatrix}
\frac{1}{8} & 0 & 0 & 0 & 0 & 0\\
0 & \frac{1}{8} & 0 & 0 & 0 & 0\\
0 & 0 & \frac{1}{8} & 0 & 0 & -r\kappa\\
0 & 0 & 0 & \frac{1}{8}+r|\kappa|^2 & 0 & 0\\
0 & 0 & 0 & 0 & \frac{1}{8}+r & 0\\
0 & 0 & -r\kappa^{*} & 0 & 0 & \frac{1}{8}
\end{pmatrix}
\end{eqnarray}
where $r=\frac{1}{4(1+|\kappa|^2)}$.\\
Therefore, criterion-1 for the density operator $\rho^{(2)}_{AB}$ reduces to
\begin{eqnarray}
&&F_{avg}(\tilde{W}^{(2)},\rho^{(2)}_{AB})-\frac{1}{8}<0
\label{avstatewitness2}
\end{eqnarray}
where $F_{avg}(\tilde{W}^{(2)},\rho^{(2)}_{AB})$ is given by
\begin{eqnarray}
F_{avg}(\tilde{W}^{(2)},\rho^{(2)}_{AB})=\frac{1-4r\alpha}{8}
\label{avgwtildequtritqubit}
\end{eqnarray}
Since the parameters $\alpha$ and $r$ lying in $[0,1]$ so $F_{avg}(\tilde{W}^{(2)},\rho^{(2)}_{AB})$ is always
less than zero in this range of parameters. Hence the state $\rho^{(2)}_{AB}$ is entangled for $0\leq \alpha \leq 1$.\\
Criterion-2 not only detect the entanglement but also provide the lower and upper bound of the measure of entanglement characterized by concurrence. Therefore, our next task is to calculate the lower and upper bound of the concurrence of the state described by the density operator $\rho^{(2)}_{AB}$. To achieve our goal, we use (\ref{qutrit-qubit2},\ref{spa1}-\ref{spa13}) to obtain the SPA-PT of $\rho^{(2)}_{AB}$, which can be expressed as
\begin{eqnarray}
\widetilde{\rho}^{(2)}_{AB}=
\begin{pmatrix}
  \tilde{t}_{11} & 0 & \tilde{t}_{13} & 0 & \tilde{t}_{15} & \tilde{t}_{16}\\
  0 & \tilde{t}_{22} & 0 & \tilde{t}_{24} & 0 & \tilde{t}_{26} \\
 \tilde{t}_{13}^{*} & 0 & \tilde{t}_{33} & 0 & \tilde{t}_{35} & 0\\
  0 & \tilde{t}_{24}^{*} & 0 & \tilde{t}_{44} & \tilde{t}_{45} & \tilde{t}_{46}\\
  \tilde{t}_{15}^{*} & 0 & \tilde{t}_{35}^{*} & \tilde{t}_{45}^{*} & \tilde{t}_{55} & 0 \\
  \tilde{t}_{16}^{*} & \tilde{t}_{26}^{*} & 0 & \tilde{t}_{46}^{*} & 0 & \tilde{t}_{66} \\
\end{pmatrix}, 0 \leq \alpha \leq 1
\end{eqnarray}
where
\begin{eqnarray}
&&\tilde{t}_{11}=\frac{54}{384}+\frac{7\alpha}{384},~~\tilde{t}_{13}=\frac{9}{128},~~\tilde{t}_{15}=-\frac{9}{128}+\frac{3\alpha}{128},\nonumber\\&&\tilde{t}_{16}=\frac{\alpha}{24},\tilde{t}_{22}=\frac{9}{64}+\frac{23\alpha}{384},~~\tilde{t}_{24}=\frac{9}{128},\nonumber\\&&\tilde{t}_{26}=-\frac{9}{128}+\frac{3\alpha}{128},\tilde{t}_{33}=\frac{77}{384}-\frac{23\alpha}{384},\tilde{t}_{35}=\frac{3}{64}+\frac{3\alpha}{128},\nonumber\\&&\tilde{t}_{44}=\frac{61}{384}-\frac{7\alpha}{384},\tilde{t}_{45}=\frac{1-\alpha}{24},\tilde{t}_{46}=\frac{3}{64}+\frac{3\alpha}{128}, \nonumber\\&&\tilde{t}_{55}=\frac{61}{384}+\frac{\alpha}{24},\tilde{t}_{66}=\frac{77}{384}-\frac{\alpha}{24}
\end{eqnarray}
The lower and upper bound of the concurrence $C(\rho^{(2)}_{AB})$ is given by
\begin{eqnarray}
\frac{1}{2}-4F_{avg}(\tilde{W}^{(2)},\rho^{(2)}_{AB})\leq C(\rho^{(2)}_{AB}) \leq F_{avg}(\rho^{(2)}_{AB},\widetilde{\rho}^{(2)}_{AB})
\label{concqutritqubit}
\end{eqnarray}
where $F_{avg}(\rho^{(2)}_{AB},\widetilde{\rho}^{(2)}_{AB})$ is given by
\begin{eqnarray}
F_{avg}(\rho^{(2)}_{AB},\widetilde{\rho}^{(2)}_{AB})= \frac{78\alpha^{2}-78\alpha+154}{768}
\label{avgrhotildequtritqubit}
\end{eqnarray}
The lower and upper bound of the concurrence $C(\rho^{(2)}_{AB})$ is shown in Figure-1.
\begin{figure}[h]
\centering
\includegraphics[scale=0.6]{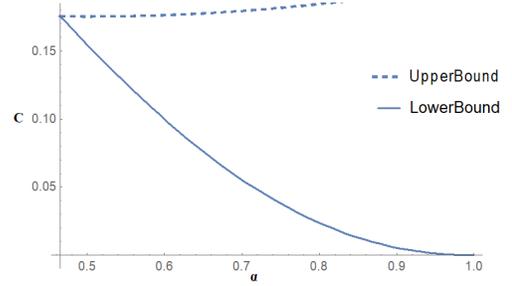}
\caption{Plot of concurrence (C) versus $\alpha$. Dotted line shows upper bound and solid line
represent lower bound of the concurrence of qutrit-qubit state described by the density
matrix $\rho^{(2)}_{AB}$ }
\end{figure}


\section{Conclusion}
To summarize, we have obtained the criteria for the detection of NPT entangled state in arbitrary dimensional Hilbert space.
The first criterion detect NPT entangled state iff the average fidelity of two quantum states described by the density matrix $\rho_{AB}$ and SPA-PT of the witness operator $\widetilde{W}$ is less than a quantity $R$. The quantity $R$ depends on (i) the dimension of the composite Hilbert space and (ii) the parameter that makes the SPA-PT of the witness operator positive semi-definite. The second criterion tells us that the given state is NPT entangled iff minimum eigenvalue of the SPA-PT of the given state is greater equal to the difference between the concurrence of the given state and the average fidelity between the given state and its SPA-PT. Since we are not able to find out the exact value of the concurrence of given state so we have derived the lower and upper bound of the concurrence. The third criterion also dealt with the detection of entanglement. Moreover, in particular, we perform SPA map on qutrit-qubit system and then explicitly calculated the elements of the SPA-PT of the qutrit-qubit system and this result will be useful to find the minimum eigenvalue of SPA-PT of the qutrit-qubit system.

\section{Acknowledgement}
Anu Kumari would like to acknowledge the financial support from CSIR.

\section{Appendix:SPA-PT of an arbitrary qutrit-qubit quantum state}
In this section, we will obtain the SPA-PT of general density matrix of qutrit-qubit quantum state.\\
To achieve our goal, let us consider an arbitrary qutrit-qubit quantum state described by the density operator in the
computational basis as
\begin{eqnarray}
\varrho_{12}=
\begin{pmatrix}
  t_{11} & t_{12} & t_{13} & t_{14} & t_{15} & t_{16} \\
  t_{12}^{*} & t_{22} & t_{23} & t_{24} & t_{25} & t_{26} \\
  t_{13}^{*} & t_{23}^{*} & t_{33} & t_{34} & t_{35} & t_{36} \\
  t_{14}^{*} & t_{24}^{*} & t_{34}^{*} & t_{44} & t_{45} & t_{46}\\
  t_{15}^{*} & t_{25}^{*} & t_{35}^{*} & t_{45}^{*} & t_{55} & t_{56}\\
  t_{16}^{*} & t_{26}^{*} & t_{36}^{*} & t_{46}^{*} & t_{56}^{*} & t_{66}
\end{pmatrix}, \sum_{i=1}^{6}t_{ii}=1
\label{qutrit-qubitstate}
\end{eqnarray}
where $(*)$ denotes the complex conjugate.\\\\
The decomposition of SPA-PT for a qutrit-qubit quantum state
$\varrho_{12}$ is given by
\begin{eqnarray}
\widetilde{\varrho_{12}}=[\frac{1}{4}(I\otimes\widetilde{T})+\frac{3}{4}(\widetilde{\Theta}\otimes\widetilde{D})]\rho_{12}
\label{spa-ptqutrit-qubit}
\end{eqnarray}
The operator $\tilde{T}[.]$ denote the SPA of partial transposition of $(.)$ and it is given by
\begin{eqnarray}
\tilde{T}[.]= \sum_{k=1}^{9}tr[M_{k}(.)]|v_{k}\rangle\langle{v_{k}}|,~~M_{k}=\frac{1}{3}|v_k^{*}\rangle\langle{v_k^{*}|}
\end{eqnarray}
Moreover, the remaining two operators $\widetilde{\Theta}$ and $\widetilde{D}$ denote the SPA of inversion map $\Theta$
and depolarization map $D$ respectively and they can be defined as
\begin{eqnarray}
\widetilde{\Theta}[.] &=& \sum_{i=1}^9 (tr[M_i(.)] \lambda_a |v_k\rangle\langle v_k|\lambda_a)\nonumber\\&=& \lambda_a \Tilde{T}[.] \lambda_a
\end{eqnarray}
\begin{eqnarray}
\widetilde{D}[.] &=& \frac{1}{4}\sum_{i=0,x,y,z}\sigma_{i}[.]\sigma_{i}
\label{spa1}
\end{eqnarray}
where $\sigma_{0}=I$, $\sigma_{i} (i=x,y,z)$ denote the Pauli matrices and
\begin{eqnarray}
&&\lambda_a=a{\lambda_a}^{(01)}+b{\lambda_a}^{(12)}+c{\lambda_a}^{(02)},~~a,b,c \in R
\nonumber\\&&{\lambda_a}^{(01)}=-i|0\rangle\langle1|+i|1\rangle\langle0|
\nonumber\\&&{\lambda_a}^{(12)}=-i|1\rangle\langle2|+i|2\rangle\langle1|
\nonumber\\&&{\lambda_a}^{(02)}=-i|0\rangle\langle2|+i|2\rangle\langle0|
\end{eqnarray}
Let $\tilde{\varrho}_{12}$ denote the SPA of partial transpose of $\varrho_{12}$. The density operator for $\tilde{\varrho}_{12}$ is given by
\begin{eqnarray}
\tilde{\varrho}_{12}=
\begin{pmatrix}
  \tilde{t}_{11} & \tilde{t}_{12} & \tilde{t}_{13} & \tilde{t}_{14} & \tilde{t}_{15} & \tilde{t}_{16} \\
  \tilde{t}_{12}^{*} & \tilde{t}_{22} & \tilde{t}_{23} & \tilde{t}_{24} & \tilde{t}_{25} & \tilde{t}_{26} \\
  \tilde{t}_{13}^{*} & \tilde{t}_{23}^{*} & \tilde{t}_{33} & \tilde{t}_{34} & \tilde{t}_{35} & t_{36} \\
  \tilde{t}_{14}^{*} & \tilde{t}_{24}^{*} & \tilde{t}_{34}^{*} & \tilde{t}_{44} & \tilde{t}_{45} & \tilde{t}_{46}\\
  \tilde{t}_{15}^{*} & \tilde{t}_{25}^{*} & \tilde{t}_{35}^{*} & \tilde{t}_{45}^{*} & \tilde{t}_{55} & \tilde{t}_{56}\\
  \tilde{t}_{16}^{*} & \tilde{t}_{26}^{*} & \tilde{t}_{36}^{*} & \tilde{t}_{46}^{*} & \tilde{t}_{56}^{*} & \tilde{t}_{66}
\end{pmatrix}, \sum_{i=1}^{6}\tilde{t}_{ii}=1
\label{qutrit-qubit2}
\end{eqnarray}
\begin{eqnarray}
&&\tilde{t}_{11}=\frac{3}{32}[(a^2+c^2)+a^2(t_{33}+t_{44})+c^2(t_{55}+t_{66})+\nonumber\\&&ac(t_{35}+{t}_{35}^{*}+t_{46}+t_{46}^{*})]
+\frac{1}{4}[\frac{2}{3}t_{11}+\frac{1}{3}t_{22}]
\label{spa1}
\end{eqnarray}
\begin{eqnarray}
\tilde{t}_{13}&=&\frac{3}{32}[bc(1+t_{55}+t_{66})-a^2(t_{13}+t_{24})-ac(t_{15}+t_{26})\nonumber\\&+&ab(t_{35}^{*}+t_{46}^{*})]
+\frac{1}{4}[\frac{2}{3}t_{13}+\frac{1}{3}t_{24}]
\label{spa2}
\end{eqnarray}
\begin{eqnarray}
\tilde{t}_{15}&=&\frac{3}{32}[-ab(1+t_{33}+t_{44})-ac(t_{13}+t_{24})-c^2(t_{15}\nonumber\\&+& t_{26})-bc(t_{35}+t_{46}]+\frac{1}{4}[\frac{2}{3}t_{15}+\frac{1}{3}t_{26}]
\label{spa3}
\end{eqnarray}
\begin{eqnarray}
\tilde{t}_{22}&=&\frac{3}{32}[(a^2+c^2)+a^2(t_{33}+t_{44})+c^2(t_{55}+t_{66})\nonumber\\&+&ac(t_{35}+{t}_{35}^{*}+t_{46}+t_{46}^{*})]+
\frac{1}{4}[\frac{1}{3}t_{11}+\frac{2}{3}t_{22}]
\label{spa4}
\end{eqnarray}
\begin{eqnarray}
\tilde{t}_{24}&=&\frac{3}{32}[bc(1+t_{55}+t_{66})-a^2(t_{13}+t_{24})-ac(t_{15}+t_{26})\nonumber\\&+&
ab(t_{35}^{*}+t_{46}^{*}]+\frac{1}{4}[\frac{1}{3}t_{13}+\frac{2}{3}t_{24}]
\label{spa5}
\end{eqnarray}
\begin{eqnarray}
\tilde{t}_{26}&=&\frac{3}{32}[-ab(1+t_{33}+t_{44})-ac(t_{13}+t_{24})-c^2(t_{15}+t_{26})\nonumber\\&-&bc(t_{35}+t_{46})]+\frac{1}{4}[\frac{1}{3}t_{15}
+\frac{2}{3}t_{26}]
\label{spa6}
\end{eqnarray}
\begin{eqnarray}
\tilde{t}_{33}&=&\frac{3}{32}[a^2+b^2+a^2(t_{11}+t_{22})+b^2(t_{55}+t_{66})\nonumber\\&-&ab(t_{15}+t_{15}^{*})-ab(t_{26}+t_{26}^{*})]+
\frac{1}{4}[\frac{2}{3}t_{33}+\nonumber\\&&\frac{1}{3}t_{44}]
\label{spa7}
\end{eqnarray}
\begin{eqnarray}
\tilde{t}_{35}&=&\frac{3}{32}[ac(1+t_{11}+t_{22})-bc(t_{15}+t_{26})+ab(t_{13}^{*}+t_{24}^{*})\nonumber\\&-&b^2(t_{35}+t_{46})]+
\frac{1}{4}[\frac{2}{3}t_{35}+\frac{1}{3}t_{46})]
\label{spa8}
\end{eqnarray}
\begin{eqnarray}
\tilde{t}_{44}&=&\frac{3}{32}[a^2+b^2+a^2(t_{11}+t_{22})+b^2(t_{55}+t_{66})-ab(t_{15}+\nonumber\\&&
t_{15}^{*})-ab(t_{26}+t_{26}^{*})]+\frac{1}{4}[\frac{1}{3}t_{33}+\frac{2}{3}t_{44}]
\label{spa9}
\end{eqnarray}
\begin{eqnarray}
\tilde{t}_{46}&=&\frac{3}{32}[ac(1+t_{11}+t_{22})-bc(t_{15}-t_{26})\nonumber\\&+&ab(t_{13}^{*}+t_{24}^{*})-b^2(t_{35}+t_{46})
+\frac{1}{4}[\frac{1}{3}t_{35}\nonumber\\&+&\frac{2}{3}t_{46}]
\label{spa10}
\end{eqnarray}
\begin{eqnarray}
\tilde{t}_{55}&=&\frac{3}{32}[(b^2+c^2)+c^2(t_{11}+t_{22})+b^2(t_{33}+t_{44})
\nonumber\\&+&bc(t_{13}+t_{13}^{*}+t_{24}+t_{24}^{*})]+\frac{1}{4}[\frac{2}{3}t_{55}+\frac{1}{3}t_{66}]
\label{spa11}
\end{eqnarray}
\begin{eqnarray}
\tilde{t}_{66}&=&\frac{3}{32}[(b^2+c^2)+c^2(t_{11}+t_{22})+b^2(t_{33}+t_{44})
\nonumber\\&+&bc(t_{13}+t_{13}^{*}+t_{24}+t_{24}^{*})]+\frac{1}{4}[\frac{1}{3}t_{55}+\frac{2}{3}t_{66}]
\label{spa12}
\end{eqnarray}
\begin{eqnarray}
&&\tilde{t}_{12}=\frac{1}{12}t_{12}^{*}, \tilde{t}_{14}=\frac{1}{12}t_{23},
\tilde{t}_{16}=\frac{1}{12}t_{25}, \tilde{t}_{23}=\frac{1}{12}t_{14},\nonumber\\&& \tilde{t}_{25}=\frac{1}{12}t_{16},
\tilde{t}_{34}=\frac{1}{12}t_{34}^{*}, \tilde{t}_{36}=\frac{1}{12}t_{45}, \tilde{t}_{45}=\frac{1}{12}t_{36},\nonumber\\&&
\tilde{t}_{56}=\frac{1}{12}t_{56}^{*}
\label{spa13}
\end{eqnarray}
The value of the parameters $a$,$b$,$c$ can be chosen in such a way that $Tr(\tilde{\varrho}_{12})=1$.






\begin{thebibliography}{90}
\bibitem{piani} M. Piani, S. Gharibian, G. Adesso, J. Calsamiglia, P.
Horodecki,and A. Winter, Phys. Rev. Lett. \textbf{106}, 220403
(2011);R. Horodecki, P. Horodecki, M. Horodecki, and K. Horodecki,
Rev. Mod. Phys. \textbf{81}, 865 (2009).
\bibitem{jozsa} R. Jozsa, and N. Linden, Proc. R. Soc. Lond. A \textbf{459}, 2011 (2003).
\bibitem{Bennett1} C. H. Bennett, G. Brassard, C. Crepeau, R. Jozsa, A. Peres, and W. K. Wootters, Phys. Rev. Lett. \textbf{70}, 1895 (1993).
\bibitem{Bennett2} C. H. Bennett and S. Wiesner, Phys. Rev. Lett. \textbf{69}, 2881 (1992).
\bibitem{pati} A. K. Pati, Phys. Rev. A \textbf{63}, 014302 (2000).
\bibitem{Gisin} C. H. Bennett, and G. Brassard, in Proceedings of the IEEE International
Conference on Computers, Systems and Signal Processing, Bangalore,
India, (IEEE, New York), 175 (1984); A. K. Ekert, Phys. Rev. Lett.
67, \textbf{661} (1991); N. Gisin, G. Ribordy, W. Tittel, and H.
Zbinden, Rev. Mod. Phys. \textbf{74}, 145 (2002).
\bibitem{guhne2} O. Guhne, G. Toth, Phys. Rep. \textbf{474}, 1 (2009).
\bibitem{augusiak} R. Augusiak, M. Demianowicz, P. Horodecki, Phys. Rev. A \textbf{77}, 030301(R) (2008).
\bibitem{lima} G. Lima, E. S. Gmez, A. Vargas, R. O. Vianna, C. Saavedra, Phys. Rev. A \textbf{82}, 012302 (2010).
\bibitem{chruscinski1} D. Chruscinski, F. A. Wudarski, Open Sys. Inf. Dyn. \textbf{18}, 387 (2011).
\bibitem{ganguly} N. Ganguly, S. Adhikari, Phys. Rev. A \textbf{80}, 032331 (2009).
\bibitem{horodecki1} P. Horodecki, R. Augusiak, M. Demianowicz, Phys. Rev. A \textbf{74}, 052323 (2006).
\bibitem{toth1} G. Toth, O. Guhne, Phys. Rev. Lett. \textbf{94}, 060501 (2005).
\bibitem{li} M. Li, S-M Fei, X. Li-Jost, and H. Fan, Phys. Rev. A \textbf{92}, 062338 (2015).
\bibitem{shen} S-Q Shen, T-R Xu, S-M Fei, X. Li-Jost, and M. Li, Phys. Rev. A \textbf{97}, 032343 (2018).
\bibitem{hou} J. Hou and X. Qi, Phys. Rev. A \textbf{81}, 062351 (2010).
\bibitem{guhne1} G. Toth, O. Guhne, Phys. Rev. A \textbf{72}, 022340 (2005).
\bibitem{toth2} G. Toth, Phys. Rev. A \textbf{71}, 010301(R) (2005).
\bibitem{peres} A. Peres, Phys. Rev. Lett. \textbf{77}, 1413 (1996).
\bibitem{horodecki2} M. Horodecki, P. Horodecki, and R. Horodecki, Phys. Lett. A \textbf{223}, 1 (1996).
\bibitem{oliver} O. Rudolph, Phys. Rev. A \textbf{67}, 032312 (2003).
\bibitem{horodecki3} P. Horoecki and A. Ekert, Phys. Rev. Lett. \textbf{89},
127902-1(2002).
\bibitem{bae} J. Bae, Rep. Prog. Phys. \textbf{80}, 104001 (2017).
\bibitem{adhikari1} S. Adhikari, Phys. Rev. A \textbf{97}, 042344 (2018).
\bibitem{adhikari2} S. Adhikari, Eur. Phys. Lett. \textbf{124}, 40006 (2018).
\bibitem{lim1} H-T. Lim, Y-S. Kim, Y-S. Ra, J. Bae, and Y-H. Kim, Phys. Rev. A \textbf{86}, 042334 (2012).
\bibitem{kwong} C. J. Kwong, S. Felicetti, L. C. Kwek, J. Bae, quant-ph/arXiv1606.00427.
\bibitem{mintert1} F. Mintert, M. Kus, A. Buchleitner, Phys. Rev. Lett. \textbf{92}, 167902-1 (2004);
\bibitem{Lasserre} J. B. Lasserre, IEEE Trans. on Automatic Control \textbf{40}, 1500 (1995);
\bibitem{mintert2} F. Mintert, Phys. Rev. A \textbf{75}, 052302 (2007);
\bibitem{rana} S. Rana, Phys. Rev. A \textbf{87}, 054301 (2013);
\bibitem{connor} K. M. O. Connor, and W. K. Wootters, Phys. Rev. A \textbf{63}, 052302 (2001);
\bibitem{bruss} D. Bruss, and C. Macchiavello, Found. Phys. \textbf{33}, 1617 (2003);
\bibitem{verstraete1} F. Verstraete, K. Audenaert, and B. D. Moor, Phys. Rev. A \textbf{64}, 012316 (2001);
\bibitem{munro} W. J. Munro, D. F. V. James, A. G. White, and P. G. Kwiat, Phys. Rev. A \textbf{64}, 030302 (2001);
\bibitem{ishizaka} S. Ishizaka, and T. Hiroshima, Phys. Rev. A \textbf{62}, 022310 (2000);
\bibitem{ali} M. Ali, quant-ph/arXiv:1905.05462;
\bibitem{karpat} G. Karpat, and Z. Gedik, Phys. Lett. A \textbf{375}, 4166 (2011);




























\end{thebibliography}
\end{document}